\documentclass[conference]{IEEEtran}
\usepackage{fancyhdr}
\IEEEoverridecommandlockouts
\usepackage{cite}
\usepackage{amsmath,amssymb,amsfonts}
\usepackage{bm}
\usepackage{graphicx}
\usepackage{balance}
\usepackage{textcomp}
\usepackage{xcolor}
\usepackage{color}
\def\BibTeX{{\rm B\kern-.05em{\sc i\kern-.025em b}\kern-.08em
    T\kern-.1667em\lower.7ex\hbox{E}\kern-.125emX}}
\usepackage{tikz}
\newcommand*{\circled}[1]{\lower.7ex\hbox{\tikz\draw (0pt, 0pt)%
    circle (.5em) node {\makebox[1em][c]{\small #1}};}}

\makeatletter
\newif\if@restonecol
\makeatother

\usepackage[linesnumbered,ruled,vlined,noend]{algorithm2e}
\usepackage{algpseudocode}
\usepackage{amsmath}
\usepackage{xcolor}

\SetCommentSty{mycommfont} 
\usepackage{setspace}
\usepackage{float}
\usepackage{url}
\usepackage[normalem]{ulem} 

\let\oldnl\nl
\newcommand{\nonl}{\renewcommand{\nl}{\let\nl\oldnl}}

\newcommand{\para}[1]{\noindent \textbf{#1 }}

\usepackage{subcaption} 
\newcommand\systemName{GraphPi}
\newcommand\algorithmName{2-cycle based automorphism elimination}

\begin{document}

\title{\systemName{}: High Performance Graph Pattern Matching through Effective Redundancy Elimination}

\author{
{\rm Tianhui Shi},
{\rm Mingshu Zhai},
{\rm Yi Xu},
{\rm Jidong Zhai} \\
Tsinghua University \\
\{sth19, dms18, xuyi18\}@mails.tsinghua.edu.cn,
zhaijidong@tsinghua.edu.cn
} 

\maketitle

\begin{abstract}
Graph pattern matching, which aims to discover structural patterns in graphs, is considered one of the most fundamental graph mining problems in many real applications. Despite previous efforts, existing systems face two main challenges. First, inherent symmetry existing in patterns can introduce a large amount of redundant computation. Second, different matching orders for a pattern have significant performance differences and are quite hard to predict. When these factors are mixed, this problem becomes extremely complicated. High efficient pattern matching remains an open problem currently.


To address these challenges, we propose \systemName{}, a high performance distributed pattern matching system. \systemName{} utilizes a new algorithm based on 2-cycles in group theory to generate multiple sets of asymmetric restrictions, where each set can eliminate redundant computation completely.
We further design an accurate performance model to determine the optimal matching order and asymmetric restriction set for efficient pattern matching. We evaluate \systemName{} on Tianhe-2A supercomputer. Results show that \systemName{} outperforms the state-of-the-art system, by up to 105$\times$ for 6 real-world graph datasets on a single node. We also scale \systemName{} to 1,024 computing nodes (24,576 cores).
\end{abstract}


\section{Introduction}
Graph data and algorithms are widely used in many fields, such as social networks\cite{fan2013diversified}, bioinformatics\cite{alon2008biomolecular}, and fraud detection\cite{choudhury2015selectivity}. With the increasing amount of graph data, processing and analyzing graphs with high performance become more and more critical. Graph analysis problems can be mainly classified into two types: graph computation and graph mining.
Graph computation problems have been extensively studied, and many efficient graph processing systems have been proposed\cite{zhu2015gridgraph,zhu2016gemini,chen2019powerlyra,gonzalez2012powergraph,gonzalez2014graphx,zheng2015flashgraph,roy2013x,roy2015chaos}.
On the other hand, efficient and scalable graph mining algorithms, which are widely used to discover complex structural patterns in graphs, are extremely challenging to design. As the most typical and common graph mining problem, pattern matching is well known to be NP-complete. With the increase of graph data scale, the number of potential pattern instances may increase exponentially, resulting in an exponential increase in searching space, computation, and intermediate data. The state-of-the-art graph pattern matching system\cite{mawhirter2019graphzero} needs several hours or even several days to mine a pattern with a size of 6 on an unlabeled graph with millions of edges\cite{backstrom2006group}. 


Recently, researchers have proposed several general-purpose graph mining systems~\cite{teixeira2015arabesque,wang2018rstream,yan2017g,chen2018g}, such as Arabesque and RStream,  which provide high-level abstractions and flexible programming models to express complex graph mining algorithms. 
Arabesque\cite{teixeira2015arabesque} is a distributed graph mining system that uses a filter-process programming model to simplify the development of scalable graph mining algorithms. However, it suffers from large startup and communication overhead and a large amount of intermediate data. RStream\cite{wang2018rstream} leverages persistent storage to store intermediate data and implements relational algebra efficiently with tuple streaming and outperforms several state-of-the-art distributed mining systems. Although these general-purpose systems provide comprehensive support for the development of a series of mining algorithms, they have relatively poor performance and exponential intermediate data for storing candidates. For example, RStream generates about 1.2TB intermediate data to count 4-motif on the MiCo graph with 1 million edges\cite{mawhirter2019automine}.

Recently, specialized systems have been developed for pattern matching\cite{reza2018prunejuice,mawhirter2019graphzero,bhattarai2019ceci,serafini2017qfrag}, approximate pattern mining\cite{ahmed2014graph,iyer2018asap}, and frequent subgraph mining (FSM)\cite{abdelhamid2016scalemine,talukder2016distributed}. ASAP\cite{iyer2018asap} is a distributed approximate pattern matching system for estimating the count of embeddings (instances of the input pattern). It allows users to make a trade-off between the result accuracy and latency. Although ASAP shows outstanding scalability, it is not applicable in some situations. For example, ASAP fails to generate relatively accurate estimation by sampling if there are very few embeddings in the graph, which are common for some real graph mining applications. 

Algorithms in these specialized pattern matching systems can be described with nested loops, and AutoMine\cite{mawhirter2019automine} and GraphZero\cite{mawhirter2019graphzero} represent relatively good performance in such systems. Observing that even a single-thread program outperforms general-purpose graph mining systems like RStream in the case of triangle counting, AutoMine generates efficient C++ code for enumerating all embeddings on a graph. A pattern has a large number of orders in which each vertex of a pattern is searched in these pattern matching algorithms, which are called schedules, and the performance of different schedules varies significantly. To address this problem, AutoMine proposes a performance estimation model to select a schedule with relatively high performance. Based on AutoMine, GraphZero further generates a set of restrictions to break symmetry in a pattern.

Despite previous efforts, there are two main limitations in current specialized pattern matching systems. The first one is that embeddings are repeatedly identified many times due to the symmetry of a pattern, which results in a large amount of redundant computation. These embeddings, which contain the same edges and vertices, are called automorphisms. GraphZero partially solves this problem by using restrictions to eliminate automorphisms, but it does not consider the influence of different restrictions on performance. Second, the performance of using different schedules of a pattern varies significantly, and it is challenging for pattern matching systems to select the optimal one, especially when patterns are large and complex. Our results (Figure~\ref{fig:schedules_performance}) show that the optimal schedule is 64.5$\times$ faster than the schedule selected by GraphZero for a pattern with 6 vertices. These two problems generally exist in most of current pattern matching systems. 



To address these challenges, we design \systemName{}, a high-performance distributed pattern matching system with four novel components: 
1) a \algorithmName{} algorithm to generate multiple sets of restrictions to effectively break the symmetry of patterns, 2) a 2-phase computation-avoid schedule generator to eliminate inefficient schedules, 3) an accurate performance prediction model to determine the optimal combination of schedule and restriction set, and 4) an optimization technique with the Inclusion-Exclusion Principle when only counting embeddings.

For an arbitrary pattern input by users, \systemName{} can automatically generate restrictions in the preprocessing stage and apply these restrictions when running pattern matching to eliminate all redundant computation caused by automorphisms. For the same pattern, there may be many different restriction sets, and each of them is able to eliminate all redundant computation. However, the performance of the pattern matching process may differ by several times when applying different restrictions. Since schedules and restrictions both have significant impacts on system performance, we model the performance of the pattern matching algorithm with different combinations of schedules and restrictions in the preprocessing stage, and choose the best one to run.

Moreover, if an application only counts the number of embeddings instead of listing them, we use Inclusion-Exclusion Principle to count the number of embeddings efficiently. We also leverage a fine-grained task partitioning technique and a work-stealing algorithm to implement a distributed pattern matching system with an OpenMP/MPI hybrid programming model.


To summarize, we make the following contributions:
\begin{itemize}
\item We propose a \algorithmName{} algorithm to reduce the number of automorphisms to one for an arbitrary pattern (Section~\ref{automorphisms_elimination}). To our best knowledge, it is the first algorithm that generates multiple different sets of restrictions for nested-loop-based graph mining algorithms.
\item We propose a 2-phase computation-avoid schedule generator to eliminate inefficient schedules (Section~\ref{schedule_generation}).
\item By building an accurate performance model, we can select the optimal  combinations of schedules and restrictions from thousands of candidates to execute the pattern matching algorithm (Section~\ref{performance_prediction}).
\item In the case of counting embeddings, we further propose a method based on the Inclusion-Exclusion Principle to accelerate the pattern matching algorithm (Section~\ref{IEP}).
\end{itemize}

Evaluation results on a large variety of real-world graph datasets and patterns show that \systemName{} outperforms the state-of-the-art pattern matching system by several orders of magnitude. Specifically, \systemName{} is up to 105$\times$ faster than GraphZero  and 154$\times$ faster than Fractal\cite{dias2019fractal} running on the same single node. After using the Inclusion-Exclusion Principle (IEP) for counting the number of embeddings, \systemName{}’s performance can be further improved by up to 1110$\times$ compared with that without using IEP. Currently, \systemName{} can scale up to 1,024 computing nodes (24,576 computing cores). \systemName{} is available on the website.~\footnote{\textit{\url{https://github.com/thu-pacman/GraphPi}}}
\section{Background and Motivation}
\label{sec:background}

\begin{figure}[h!]
\centerline{\includegraphics[width=0.48\textwidth]{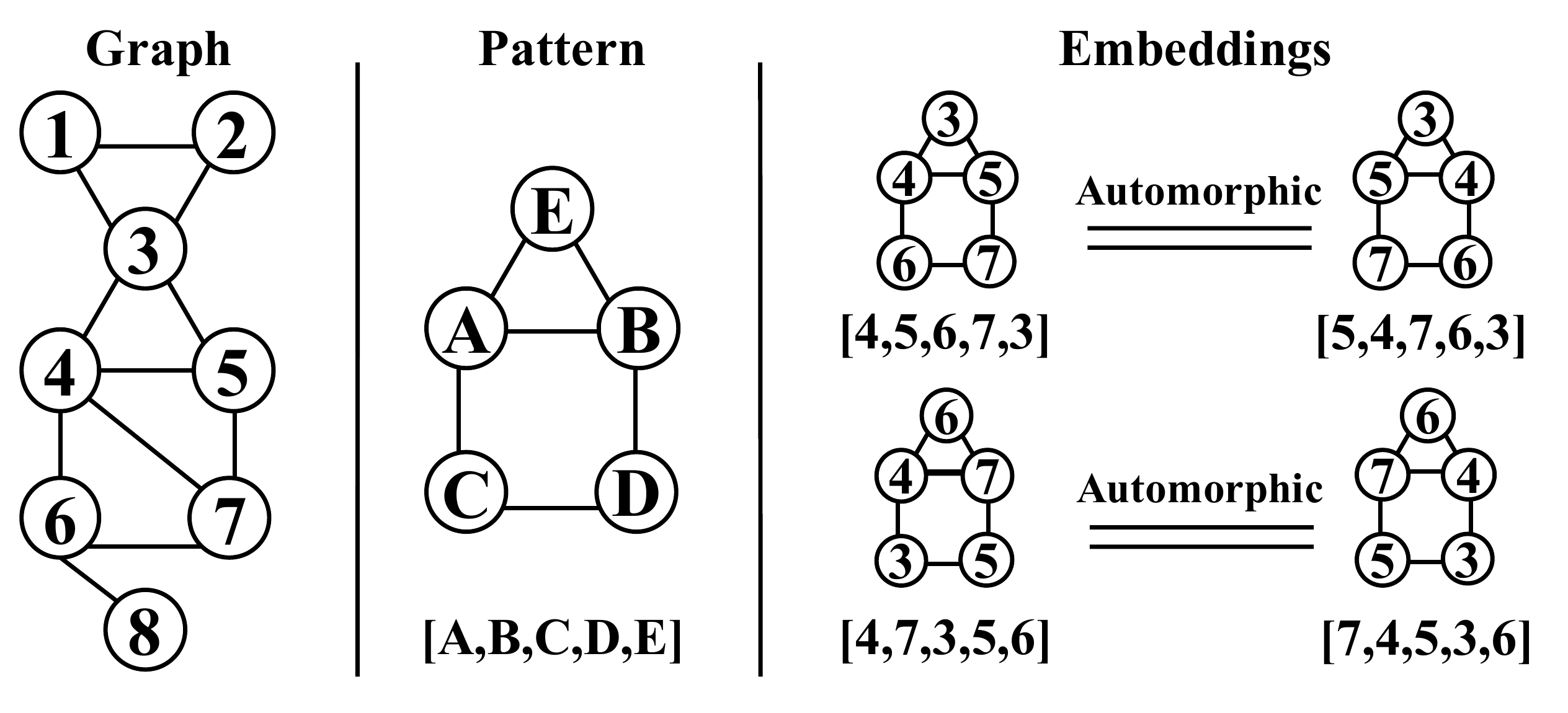}}
\caption{An example of pattern matching. The numeric letters denote different vertices in the input graph. The capital letters denote different vertices in the pattern. The numeric numbers in a square bracket denote a one-to-one correspondence (a bijective function). For example, [4,5,6,7,3] denotes the function $id$ where $id(A)=4$, $id(B)=5$, etc.}
\label{fig:pattern_matching_example}
\end{figure}

\subsection{Problem Definition} 
\label{problem_definition}

A graph $G$ is defined as $(V,E)$ where $V$ is the vertex set and $E\subseteq(V\times V)$ is the edge set. A pattern $G_p = (V_p,E_p)$ is \textbf{isomorphic} to a subgraph $G_s = (V_s,E_s)$ of the data graph if and only if there is a bijective function (a one-to-one correspondence) $id : V_p \to V_s$ such that $\forall (u_i,u_j) \in E_p$, $\big(id(u_i),id(u_j)\big) \in E_s$. For the convenience of discussion, all patterns and data graphs are assumed to be undirected and unlabeled graphs, although all methods proposed in this paper can be easily extended to directed and labeled graphs.

The graph pattern matching problem is to find all distinct subgraphs (called \textbf{embeddings}) that are isomorphic with a given pattern for a data graph. 
For better understanding, a pattern and an embedding can be regarded as a template and an instance, respectively. Figure \ref{fig:pattern_matching_example} is an example of the pattern matching problem. There are 8 and 5 vertices in the data graph and the pattern respectively. We can find 4 distinct one-to-one correspondences ([4,5,6,7,3], [5,4,7,6,3], etc.) satisfying the definition of ``isomorphic". Therefore, there are 4 embeddings of the pattern in the data graph.

One subgraph can be identified as embeddings for several times. The embeddings \textit{containing the same edges and vertices} are called \textbf{automorphisms}. For example, although the 2 embeddings at the top left and top right corners in Figure \ref{fig:pattern_matching_example} have different one-to-one correspondences, they are automorphisms due to the same edges and vertices they contain.

\begin{figure}[h!]
\centerline{\includegraphics[width=0.48\textwidth]{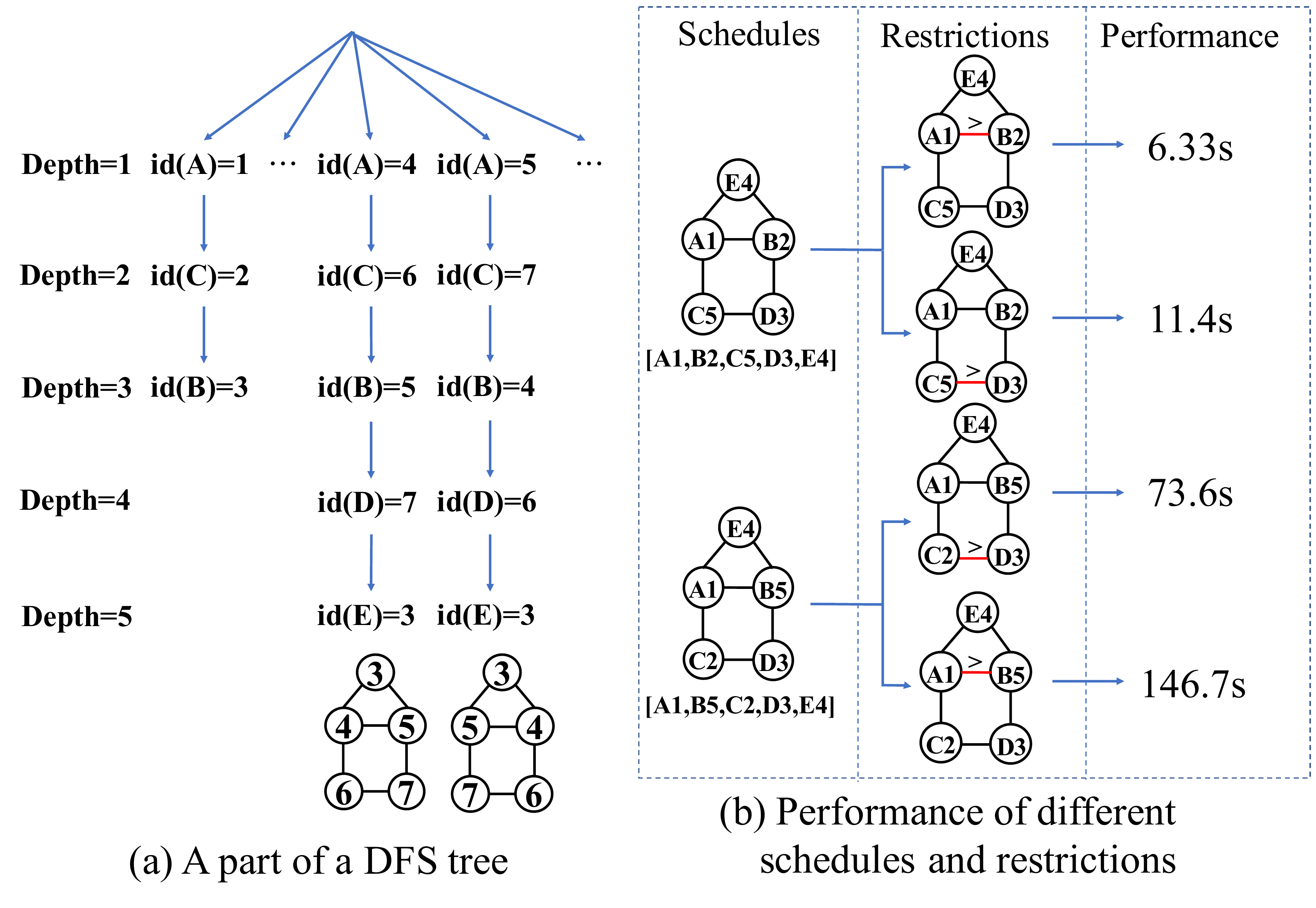}}
\caption{An example of schedules and restrictions. The schedule used in (a) is $A, C, B, D, E$. The number behind a capital letter in (b) denotes the position of this vertex in the schedule (e.g., ``C2" means the vertex C is the second searched vertex in this schedule). The greater-than sign (``$>$") above an edge denotes the restriction between two vertices of this edge (e.g., the ``$>$" above the edge between A1 and B5 denotes $id(A)>id(B)$). (b) is evaluated on the Patents graph\cite{leskovec2005graphs}.}
\label{fig:schedule_restrictions_example}
\end{figure}

\subsection{Schedules and Restrictions}




For the graph pattern matching algorithm, a \textbf{schedule} represents an order in which each vertex of a pattern is searched. For example, an order of $A, C, B, D, E$ is used to find all embeddings of the pattern in the data graph used in Figure \ref{fig:pattern_matching_example}. For a given pattern, there are usually a number of candidate schedules to perform the search. By regarding a pattern matching algorithm as a depth-first search (DFS) algorithm, Figure \ref{fig:schedule_restrictions_example}(a) shows a part of the DFS tree for this example. The deepest leaf nodes ($depth=5$) in the DFS tree represent embeddings. Note that the 2 embeddings in Figure \ref{fig:schedule_restrictions_example}(a) are automorphisms.

The number of automorphisms depends on the input pattern, and it will explode with the number of vertices in the pattern. For a 7-clique pattern (a complete graph with 7 vertices), each embedding has 5,040 automorphisms. A large number of automorphisms lead to huge redundant computation. Therefore, it is a common goal to identify each embedding exactly once in all pattern matching systems to avoid redundant computation. 

To deal with automorphisms, a technique named restriction is applied in the pattern matching algorithms. A \textbf{restriction} is a restricted condition of relative magnitudes of two vertices in a pattern (e.g., $id(A)>id(B)$, where $id$ is the one-to-one correspondence mentioned in Section~\ref{problem_definition}). By using the ordering of symmetry breaking technique\cite{grochow2007network} and the concept of the neighborhood equivalence class\cite{han2013turboiso}, the first embedding in Figure \ref{fig:schedule_restrictions_example}(a) can be eliminated with a restriction of $id(A) > id(B)$. This is because $id(A)=4<id(B)=5$ in the first embedding. From the perspective of the DFS tree, the subtree with ``$id(B)=5$" as a root node is pruned with this restriction. Therefore, the deepest leaf node of this embedding does not need to be searched.

For complex patterns, multiple restrictions are required to eliminate all redundant computation, that is, a set of restrictions is required. However, the method with neighborhood equivalence class is only effective for a part of patterns. Recently, GraphZero\cite{mawhirter2019graphzero} partially solves this problem by reducing the number of automorphisms of any pattern to 1, but it can only generate one set of restrictions and does not consider the performance difference of different sets of restrictions in pattern matching. In fact, for the same pattern, there are many different sets of restrictions that can completely eliminate redundant computation. For instance, we can use a restriction $id(C)>id(D)$ instead of $id(A) > id(B)$ to eliminate automorphisms, and the performance of a pattern matching algorithm applying the former is much higher than that applying the latter with the schedule of $A, C, B, D, E$ in Figure \ref{fig:schedule_restrictions_example}.

\begin{figure*}[htp!]
\centerline{\includegraphics[width=0.95\textwidth]{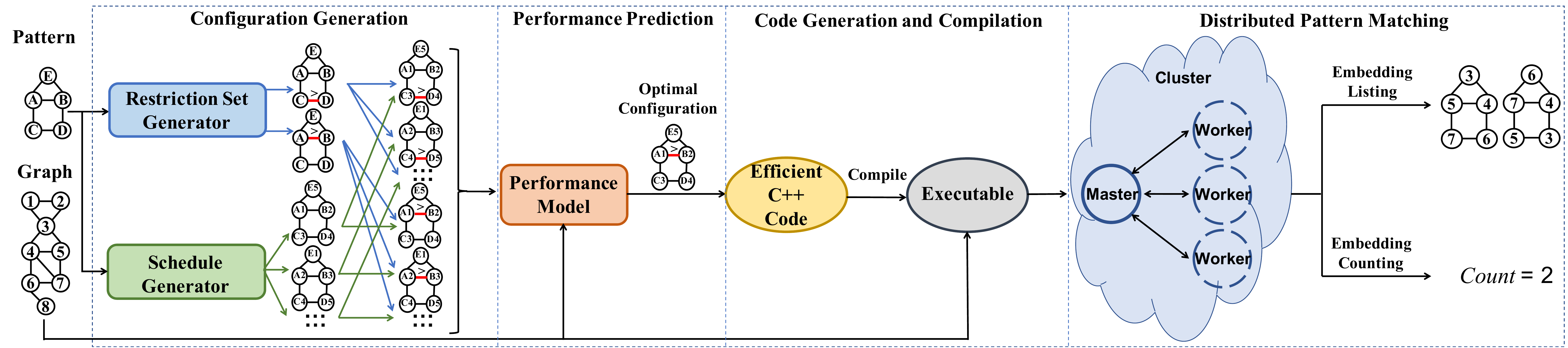}}
\caption{Overview of \systemName{}.}
\label{fig:overview}
\end{figure*}

\subsection{Challenges of Pattern Matching} \label{schedules_restrictions}

The performance of different combinations of schedules and restrictions varies significantly. For a given pattern and data graph, different schedules correspond to different searching space, that is, different DFS trees. Although the numbers of embeddings are the same for different schedules, the sizes of different DFS trees differ greatly. Applying different sets of restrictions for the same schedule can be regarded as pruning the DFS tree at different locations, which can lead to different optimization and performance. After combining schedules and a set of constraints, the performance gap between different combinations will further widen. We evaluate the performance of different combinations shown in Figure \ref{fig:schedule_restrictions_example}(b) with GraphZero on the Patents graph\cite{leskovec2005graphs}. Experimental results show that the best combination of schedules and restriction sets is up to $23.2\times$ faster than the worst one.

However, the number of combinations is very large, and it is very hard to predict the execution time of the pattern matching algorithm. There are $n!$ different schedules for a pattern with $n$ vertices. After combining with different sets of restrictions, the number of possible combinations will explode. Since the distributions of data in different graphs are not similar, we cannot do some preprocessing to select the optimal schedule and restrictions of a pattern and use a combination for one pattern in all graphs. Moreover, the sizes of DFS trees of different combinations are unknown before running the pattern matching algorithm. Therefore, it is necessary but challenging to predict the performance with low overhead and high accuracy for every combination of schedules and sets of restrictions for efficient pattern matching. 

\section{Overview of \systemName{}}

In this work, we design \systemName{}, a fast and scalable graph pattern matching system. Figure \ref{fig:overview} shows the overall architecture of \systemName{}. \systemName{} mainly explores how to select a better schedule and a set of restrictions through the information provided by the input pattern and data graph to accelerate the pattern matching process. \systemName{} consists of four main components: configuration generation, performance prediction, code generation and compilation, and distributed pattern matching.

In order to eliminate all redundant computation caused by automorphisms, different sets of restrictions of the input pattern are generated by a restriction generator. The schedule generator can generate all efficient schedules for the input pattern. A performance model predicts the relative performance of all configurations, which are combinations of restriction sets and schedules. After selecting the optimal configuration through the performance model, \systemName{} uses the pattern matching algorithm and the code generation method proposed by AutoMine to generate efficient C++ code with this configuration and compiles these programs into an executable file. Then \systemName{} runs the distributed pattern matching algorithm with the input graph to find all embeddings. If users only want the number of embeddings instead of listing embeddings, we propose using the Inclusion-Exclusion Principle in the pattern matching algorithm to further improve the performance in \systemName{}.

The APIs provided by \systemName{} are simple and easy to use. Users only need to input a pattern and a data graph in the form of adjacency lists to run \systemName{}.

\section{Methodology of \systemName{}}
This section provides a detailed description of approaches in \systemName{}. Our goal is to reduce the count of automorphisms to one and eliminate all redundant computation. To achieve that, we first introduce a novel restriction generation algorithm based on permutation groups. To our best knowledge, this is the first algorithm that generates a complete set of restrictions, which can provide more choices and optimization opportunities for graph pattern matching algorithms. 
In Subsection~\ref{schedule_generation}, we elaborate the approach how we generate efficient schedules.
In Subsection~\ref{performance_prediction}, we introduce a concept of \textbf{configuration}, which is a combination of a schedule and a set of restrictions. 
To select the optimal configuration, we design an accurate performance prediction model to estimate the relative performance of the pattern matching algorithm for each configuration.  Finally, we also propose an embedding counting method based on the Inclusion-Exclusion Principle.


\subsection{2-Cycle Based Automorphism Elimination}
\label{automorphisms_elimination}

\begin{figure}
\centerline{\includegraphics[width=0.48\textwidth]{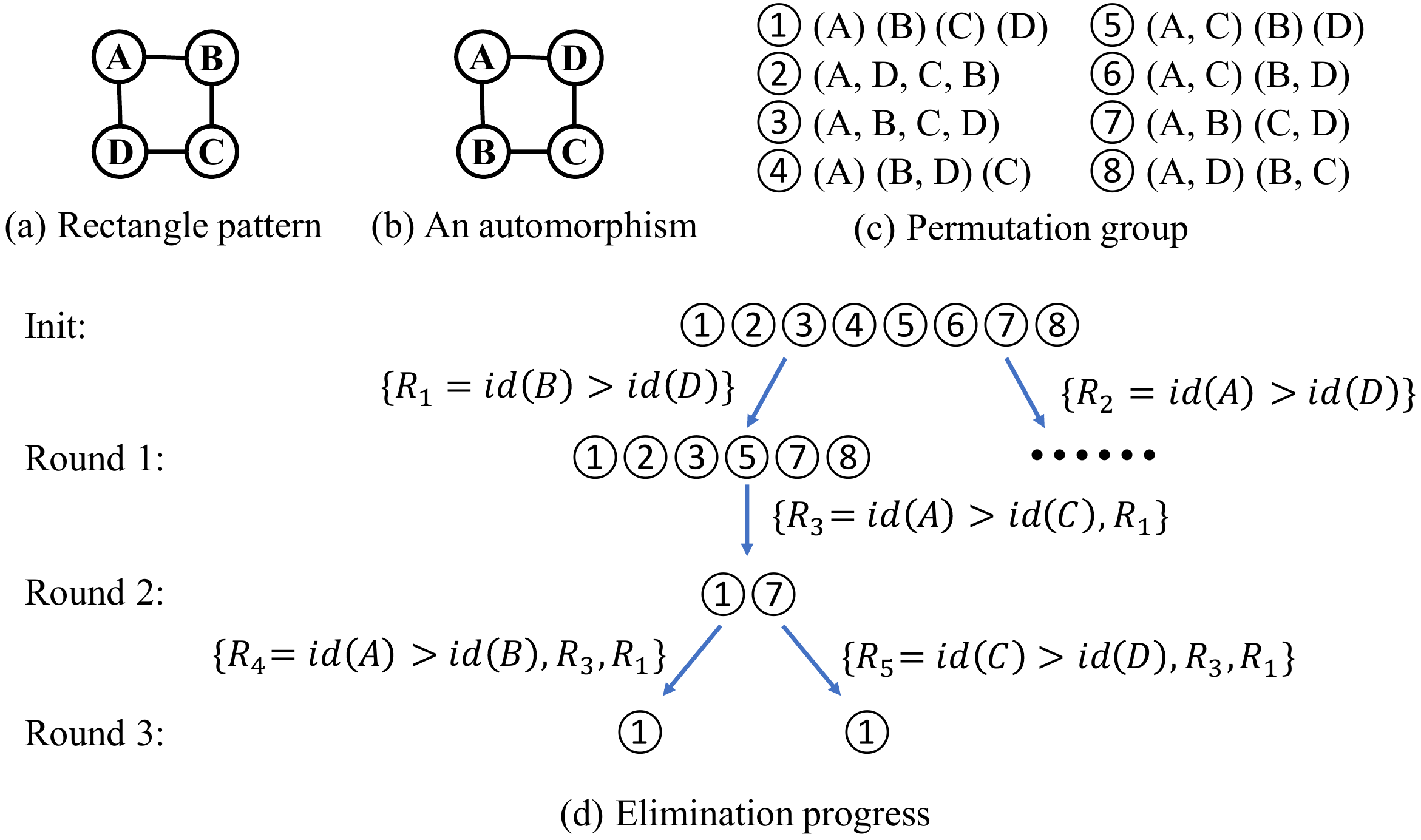}}
\caption{An example of Algorithm \ref{alg:restrictions_generation}. 
(c) is the permutation group formed of all the automorphisms of (a) and each permutation is written as a product of disjoint cycles. (d) is an example of using Algorithm 1 to eliminate all permutations in (c) except the identity permutation.}
\label{fig:permutation_group}
\end{figure}


In \systemName{}, we use the concept of permutation group \cite{wielandt2014finite} to formalize our problem. Each automorphism of a pattern can be defined as a permutation function $p : V_p \to V_p$ such that $\forall (u_i,u_j) \in E_p$, $(p(u_i),p(u_j)) \in E_p$, where $V_p$ and $E_p$ are the vertex set and edge set of a pattern. All automorphisms of a pattern form a permutation group. For example, Figure \ref{fig:permutation_group}(a) is a rectangle pattern, and Figure \ref{fig:permutation_group}(b) is one of the automorphisms of this pattern. The permutation of this automorphism is 
$\bigl(\begin{smallmatrix}
  A & B & C & D \\
  A & D & C & B
\end{smallmatrix}\bigr)$, and it can be also rewritten as a product of disjoint cycles $(A) (B, D) (C)$, where A and C are fixed (also called 1-cycle), and B and D are exchangeable (also called 2-cycle). Without loss of generality, any k-cycle ($k>1$) can be written as a product of 2-cycles. For example, a 4-cycle $(A, B, C, D)$ can be decomposed into $(A, D) (A, C) (A, B)$, which means that we first exchange A and D, then exchange A and C, and finally exchange A and B. Therefore, \textit{2-cycle exchanges are the most essential elements in permutations}. Figure \ref{fig:permutation_group}(c) is a permutation group formed by all automorphisms of this rectangle pattern. Since any k-cycle $(k>1)$ can be decomposed into 2-cycles, any permutation in a permutation group can be  rewritten as a product of 1-cycles and 2-cycles.

According to previous studies, we know that automorphisms of embeddings can be eliminated when some restrictions are applied, but the challenge is that which restrictions should be applied. On one hand, if insufficient restrictions are applied, redundant computation cannot be completely eliminated. On the other hand, if too many restrictions are applied, some embeddings will be mistakenly eliminated. In fact, the root cause of repeated embeddings is that a pattern has multiple automorphisms. In this work, we innovatively exploit the property of permutation group to effectively eliminate automorphisms of a pattern. Our core idea is to break the symmetries of permutations by applying restrictions on the essential elements of 2-cycles.

We use a concrete example to illustrate the basic idea of our approach. Suppose there is an embedding $e_1(x_1, x_2, x_3, x_4)$ \big($id(A)$=$x_1$, $id(B)$=$x_2$, etc.\big) of the rectangle pattern in a given graph. Due to the existence of the permutation $(A)(B, D)(C)$ in Figure \ref{fig:permutation_group}(c), there must be another embedding $e_2(x_1, x_4, x_3, x_2)$, which is one of the automorphisms of $e_1$. Since $id(B)=x_2,id(D)=x_4$ for $e_1$ and $id(B)=x_4,id(B)=x_2$ for $e_2$, no matter $x_2>x_4$ or $x_4>x_2$, one  of $e_1$ and $e_2$ can be definitely eliminated when applying a restriction $id(B)>id(D)$, which inspires us to apply restrictions on 2-cycles like $(B, D)$. From the perspective of permutation group, the elimination of $e_2$ can be regarded as the elimination of the permutation $(A)(B, D)(C)$. Therefore, to eliminate all redundant computation caused by automorphisms, we need to eliminate permutations with restrictions on 2-cycles until only an identity permutation \big(i.e., $(A)(B)(C)(D)$\big) remains in a permutation group.

As shown in Figure \ref{fig:permutation_group}(d), not all 2-cycles need to be used in one set of restrictions, and therefore the selection of different 2-cycles results in different sets of restrictions. As a restriction associated with a 2-cycle can not only influence those permutations with the same 2-cycles, it can also influence other permutations, we find that even in a permutation with no 2-cycles written as a product of \textit{disjoint} cycles, it can be still eliminated by restrictions applied for other permutations, which will be discussed later.

\begin{algorithm}[ht]
  \SetAlgoNoLine 
  \DontPrintSemicolon
  \caption{\algorithmName{}}
  \label{alg:restrictions_generation}
  \KwIn{$pattern$: the pattern}
  \KwOut{$res\_sets$: sets of restrictions}
  \SetKwFunction{FMain}{res\_set\_generation}
  \SetKwProg{Fn}{Function}{:}{}
  \Fn{\FMain{$pattern$}}{
      $auts\leftarrow$ all the automorphisms of $pattern$\;
      $pg\leftarrow$ the permutation group formed of $auts$\;
      $res\_sets \leftarrow$ generate($pg$, $\varnothing$)\;
      \KwRet $res\_sets$
  }
  
  \nonl\;
  \SetKwFunction{FRecursive}{generate}
  \SetKwProg{Fn}{Function}{:}{}
  \Fn{\FRecursive{$pg$, $res\_set$}}{
  \label{func:generate}
    \If{$pg$.\textup{size} $> 1$}
    {
        $sets \leftarrow \varnothing$\;
        \For{$perm \in pg$}
        {
            \For{$vertex \in perm$}
            {
                \If(\;\tcp*[h]{a 2-cycle is found}){$vertex = perm[perm[vertex]]$}
                {
                    $new\_set \leftarrow$ \;
                    \nonl $res\_set$ $\cup$ pair($vertex, p[vertex]$)\;
                    $remaining\_pg \leftarrow \varnothing$\;
                    \For{$p \in pg$}
                    {
                        \If{\textup{no\_conflict}($p$, $new\_set$)}
                        {
                            $remaining\_pg$.add($p$)\;
                        }
                    }
                    $sets \leftarrow sets \cup generate(remaining\_pg,new\_set)$
                }
            }
        }
        \KwRet $sets$\;
      }
    \Else(\tcp*[h]{only the identity permutation})
    {
        \If{\textup{validate}($res\_set$)}{\KwRet $res\_set$}
        \Else{\KwRet $\varnothing$}
    }
  }
  
  \nonl\;
  \SetKwFunction{FCheck}{no\_conflict}
  \SetKwProg{Fn}{Function}{:}{}
  \Fn{\FCheck{$perm$, $res\_set$}}{
  \label{func:no_conflict}
    $g \leftarrow$ an empty directed graph\;
    \For{$res \in res\_set$}
    {
        $g$.add\_dir\_edge($res$.first, $res$.second)\;
        $g$.add\_dir\_edge($perm[res$.first$]$, $perm[res$.second$]$)\;
    }
    \KwRet $g$.\textup{acyclic()} \;
  }
\end{algorithm}

Algorithm \ref{alg:restrictions_generation} presents our \algorithmName{} algorithm. It takes an arbitrary pattern as input and outputs multiple sets of restrictions, and each of them can eliminate all redundant computation. We first generate all the automorphisms and the corresponding permutation group (lines 2$\sim$3). Then, we call a recursive function $generate$ (line 4). If there are permutations other than an identity permutation in the group, more restrictions need to be added to eliminate them (line 7). For each permutation in the group, if we succeed in finding a 2-cycle in the permutation, we append a partial order restriction between the two vertices of the 2-cycle (lines 9$\sim$12). Next, we use a new set of restrictions to eliminate permutations (lines 13$\sim$16). In order to eliminate the remaining permutations, we generate more restrictions by calling the function of $generate$ recursively (line 17). When there is only one identity permutation in the group, we verify the current set of restrictions 
by calling the $validate(res\_set)$ function (lines 19$\sim$20). Assuming that the pattern has $n$ vertices, we run a pattern matching algorithm with the input set of restrictions (i.e., $res\_set$) on an n-vertex complete graph and rerun it without restrictions. The set of restrictions is correct if $ans_{with} = ans_{without}/automorphisms\_count$, where $ans_{with}$ is the number of embeddings found during the pattern matching process with the set of restrictions and $ans_{without}$ is that without restrictions (i.e., including all automorphisms).

The function of $no\_conflict$ is used to verify whether a permutation can be eliminated by the current set of restrictions. For each restriction in the set, we regard it as two directed edges and add them to an initially empty directed graph $g$. The permutation is not eliminated if and only if $g$ is a directed acyclic graph (DAG). For example, there are two restrictions \big($id(B) > id(D)$ and $id(A) > id(C)$\big) after Round 1 in Figure \ref{fig:permutation_group}(d). We assume that the permutation \circled{2} is not eliminated. After applying these two restrictions, if the pattern matching algorithm can find an embedding $e_1(x_1,x_2,x_3,x_4)$ \big($id(A)$=$x_1$, $id(B)$=$x_2$, etc.\big), then another embedding $e_2(x_4,x_1,x_2,x_3)$ can also be found due to the permutation. Since $e_1$ meets the restrictions, there must be \textbf{$x_2>x_4$} and $x_1>x_3$. Similarly, there must be $x_1>x_3$ and \textbf{$x_4>x_2$} for $e_2$. Then, we have two contradictory relations $x_2>x_4$ and $x_4>x_2$ (corresponding to a ring in the directed graph $g$). Therefore, the assumption does not hold. That is, the permutation \circled{2} is eliminated by these two restrictions.

In our evaluation, we will show that Algorithm \ref{alg:restrictions_generation} not only generates multiple different sets of restrictions but also has very low overhead. By contrast with the execution time of the pattern matching algorithm, which may take several minutes or even several hours, the overhead of Algorithm \ref{alg:restrictions_generation} can be ignored. 


\subsection{2-Phase Computation-Avoid Schedule Generation} \label{schedule_generation}
In pattern matching, a schedule represents an order in which each vertex of a pattern is searched. For a pattern with $n$ vertices, there are $n!$ possible schedules. How to select an efficient schedule is a key challenge in pattern matching. 

\begin{figure}[t]
\setlength{\belowcaptionskip}{0.5em}
\centerline{\includegraphics[width=0.48\textwidth]{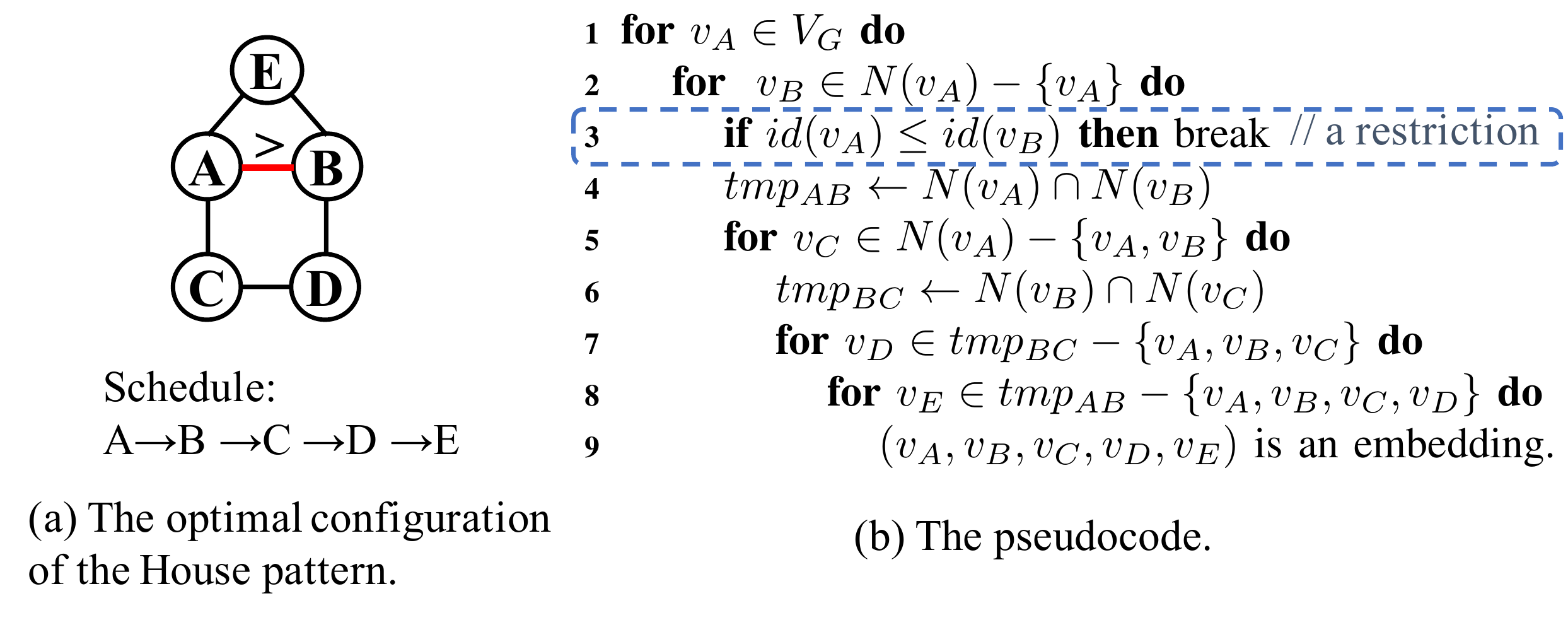}}
\caption{For a given configuration like (a), \systemName{} can generate efficient code like (b) to find all embeddings. $V_G$ is the vertex set of the input data graph. $N(v)$ returns the neighborhood of $v$ in the data graph.}
\label{fig:pseudocode_performance}
\end{figure}

Typically, an efficient pattern matching algorithm is usually implemented using recursive functions or nested loops. For example, Figure~\ref{fig:pseudocode_performance}(b) is the pseudocode of the algorithm generated by \systemName{} for the House pattern with a schedule of $A\rightarrow{} B\rightarrow{} C\rightarrow{}D\rightarrow{}E$ and a restriction of $id(A)>id(B)$. For simplicity, we temporarily ignore the statement (line 3) related to the restriction. The algorithm searches every vertex in the order specified by the schedule. In this work, we let a \textbf{candidate set} of a vertex be a set where the vertex traverses in a loop. A candidate set is either the neighborhood of a vertex or the intersection of neighborhoods of several vertices in the data graph. For example, the vertex $E$ is connected to $A$ and $B$ in Figure \ref{fig:pseudocode_performance}(a), so in the pseudocode, $v_E$ traverses through the intersection of neighborhoods of $v_A$ and $v_B$ in the data graph, that is, the candidate set of E is $N(v_A)\cap N(v_B)$.

Although there are $n!$ possible schedules for a given pattern, some of them are inefficient. Though our performance prediction module can predict the performance of the schedules generated by the schedule generator, generating efficient schedules instead of all schedules can significantly reduce the overhead of performance prediction. Based on our observation, the overhead of intersection operations is the main cost in pattern matching. Therefore, we propose a 2-phase computation-avoid schedule approach to generate efficient schedules. Our approach consists of two key phases. 

\para{Phase 1} We firstly eliminate schedules which do not satisfy that \textit{the $i$th searched vertex is directly connected to at least one of the first $i-1$ searched vertices} in the pattern. That is, for an efficient schedule, \textit{the subgraph formed by the first i searched vertices must be a connected graph}. For example, if we first search vertex C and vertex D in Figure~\ref{fig:pseudocode_performance}(a) and the third searched vertex is E, this schedule is inefficient regardless of what the search order of the remaining two vertices is, because there is no edge between E and C or E and D in the pattern. This rule can be further explained. For the third searched vertex E, since E is not connected to C or D, the candidate set of E is the entire vertex set (i.e., $V_G$). If we use vertex A instead of E as the third searched vertex, the candidate set of A will be the intersection of neighborhoods of C and D (i.e., $N(
C) \cap N(D)$). Obviously, $|V_G|$ is much larger than $|N(
C) \cap N(D)|$ in real-world graphs, which results in more computation in inner loops. So it is not an efficient schedule to choose E as the third searched vertex when C and D are the first two searched vertices.

\para{Phase 2} Suppose there are at most k vertices in a pattern which satisfy that any two of them are not directly connected. We secondly eliminate the schedules which do not satisfy that \textit{any two of the last k searched vertices are not directly connected in the pattern}. For example, the vertex D is not connected to E in Figure \ref{fig:pseudocode_performance}(a) and therefore $k=2$ for this pattern. We can see that D and E are searched in the innermost 2 loops in the pseudocode. This rule can be further explained. The last k searched vertices in a schedule are not directly connected in the pattern, which means that the candidate sets of the last k searched vertices are computed in the outermost $n-k$ loops and \textit{there are no intersection operations in the innermost k loops}. Since inner loops are executed more times than outer loops, putting intersection operations in outer loops achieves relatively less computation in total.

By generating all $n!$ possible schedules and filtering inefficient ones out with the above approach, we get all efficient schedules for further analysis.

\subsection{An Accurate Performance Prediction Model} 
\label{performance_prediction}

A main challenge in the graph pattern matching algorithm is that the performance of different schedules and a set of restrictions varies significantly for the same pattern. In this work, we use \textit{configuration} to denote a combination of a schedule and a set of restrictions. A pattern is used to indicate what kind of subgraph structures to find, while a configuration of this pattern is used to indicate how to find these subgraphs efficiently. To achieve much higher performance, we propose an accurate performance prediction module to determine the optimal configuration. 



As shown in Figure \ref{fig:pseudocode_performance}(b), the algorithm generated by \systemName{} consists of nested loops. The overhead of intersection operations is the main overhead in this algorithm, and the number of intersection operations is determined by both loops and break statements. There are three factors affecting the performance of the algorithm: the cardinality of a set where a loop traverse (e.g., line 1), the probability of breaking a loop because of not satisfying a restriction (e.g., line 3), and the overhead of intersecting two sets (e.g., line 4).

Since the algorithm uses nested loops for computation, we design a nested performance model correspondingly:
\begin{equation}\nonumber
\setlength\abovedisplayskip{7pt}
\setlength\belowdisplayskip{7pt}
    cost_i=
    \begin{cases}
        l_i \times (1-f_i) \times (c_i + \bm{cost_{i+1}}), for\; 1\leq i \leq n-1\\
        l_i \times (1-f_i) , for \; i=n
    \end{cases}
\end{equation}
where $n$ is the number of vertices of the pattern, $cost_i$ is the total cost of the $i$th loop, $l_i$ is the loop size (i.e., the cardinality of a candidate set), $f_i$ is the probability that one embedding is filtered out by a restriction, and $c_i$ is the computation overhead of the intersection operations. Since the value of $cost_i$ depends on $cost_{i+1}$, we need to calculate $cost_i$ recursively. Next, we describe how we calculate each factor.

\para{Measurement of $c_i$ and $l_i$} The neighborhood of a vertex in the data graph can be sorted in advance, so the time complexity of the intersection operations is $O(card_1+card_2)$, where $card_1$ and $card_2$ are the cardinalities of the two sets intersected. For example, there is an intersection operation $N(v_A) \cap N(v_B)$ (line 4) in the second loop in Figure \ref{fig:pseudocode_performance}(b), so $c_2 = |N(v_A)| + |N(v_B)|$. $l_i$ can be also represented by the cardinality of a set \big(e.g., for the fourth loop in line 7, $l_4 = |tmp_{AB}| = |N(v_A)\cap N(v_B)|$\big). Therefore, we can get the values of $c_i$ and $l_i$ by predicting the cardinalities of different sets.

\para{Estimation of Cardinalities} The sets whose cardinalities need to be predicted can be divided into two categories: the neighborhood of a vertex and the intersection of neighborhoods of several vertices. For the neighborhood of a vertex, its expected cardinality is $\frac{2\times|E_G|}{|V_G|}$, where $E_G$ and $V_G$ are the edge set and the vertex set of the data graph respectively. For the intersection of two neighborhoods, we use the structural information of the data graph to predict its cardinality. Suppose $x$ is the $i$th searched vertex in the pattern, and $v_x$ traverses through $N(v_y)\cap N(v_z)$ in the $i$th loop, where $x$, $y$ and $z$ are different vertices in the pattern. Based on the first phase in Section~\ref{schedule_generation}, the subgraph formed by the first $i-1$ searched vertices is a connected graph. Since $y$ and $z$ belong to the first $i-1$ searched vertices, they are connected. So that $x$, $y$ and $z$ must be in at least one ring of the subgraph formed by the first $i$ searched vertices. For example, $v_E$ traverses through $N(v_A)\cap N(v_B)$, A, B, and E form a triangle in the pattern. Therefore, we can use the number of triangles in the data graph to predict $|v_E|$, which is $\frac{tri\_cnt}{2\times |E_G|}$. We assume that the data graph is immutable so that the number of triangles ($tri\_cnt$) can be regarded as a constant value. Even if the graph is mutable, it is trivial to calculate $tri\_cnt$ incrementally. For rings that are not triangles, it may take too much time to count all rings in the graph, so we uniformly use the number of triangles for prediction. Without loss of generality, for the intersection of $n$ neighborhoods, the predicted cardinality of the intersection is $|V_G| \times p_1 \times p_2^{n-1}$, where
\vspace{0.3em}
\begin{subequations}
\setlength\abovedisplayskip{7pt}
\setlength\belowdisplayskip{7pt}
    \nonumber
    \begin{align}
    p_1  = \frac{2 \times |E_G|}{|V_G|^2} , p_2  = \frac{tri\_cnt \times |V_G|}{(2 \times |E_G|)^2} .
    \end{align}
\end{subequations}
\vspace{0.3em}
Semantically speaking, $p_1$ is the probability of any pair of vertices being neighbors \big(i.e., $P{ \left( {(a,b) \in E_G \left| a,b \in V_G\right. } \right) }$\big), and $p_2$ is the probability of any pair of vertices in a neighborhood being directly connected to each other \big(i.e., $P{ \left( {(a,b) \in E_G \left| c \in V_G, a,b \in N(c)\right. } \right) }$\big).

\para{Measurement of $f_i$} We use $f_i$ to measure the impact of restrictions on performance. For a pattern with $n$ vertices, there are $n!$ possible \textit{relative magnitudes} of $n$ vertices in an embedding (e.g., when $n=5$, they are [1,2,3,4,5], [1,2,3,5,4], [1,2,4,3,5], etc). $f_i$ is the probability that one embedding will be filtered out by the restriction in the $i$th loop. We first initialize a set of $S$ with $n!$ possible relative magnitudes. According to the order in which each restriction appears in the algorithm, we use each restriction in turn to filter out the elements in $S$ that do not satisfy the restriction. If there are several restrictions in different loops, each restriction needs to filter the elements in $S$ that are not filtered out by previous restrictions, and then we can get each value of $f_i$ by calculating the ratio of the elements filtered out. For example, since $id(A)=1<id(B)=2$ for [1,2,3,4,5], it will be filtered out by $id(A)>id(B)$. And $\frac{n!}{2}$ possibilities of relative magnitudes will be filtered out by $id(A)>id(B)$ in Figure \ref{fig:pseudocode_performance}(b), therefore, $f_1=\frac{1}{2}$. We set $f_i=0$ if there is no restriction in the $i$th loop.

Through combining above estimated parameters, we can generate an accurate performance prediction for each configuration and output a configuration with the best performance for the pattern matching algorithm.

\subsection{Counting with Inclusion-Exclusion Principle} \label{IEP}

\begin{figure}
\centerline{\includegraphics[width=0.48\textwidth]{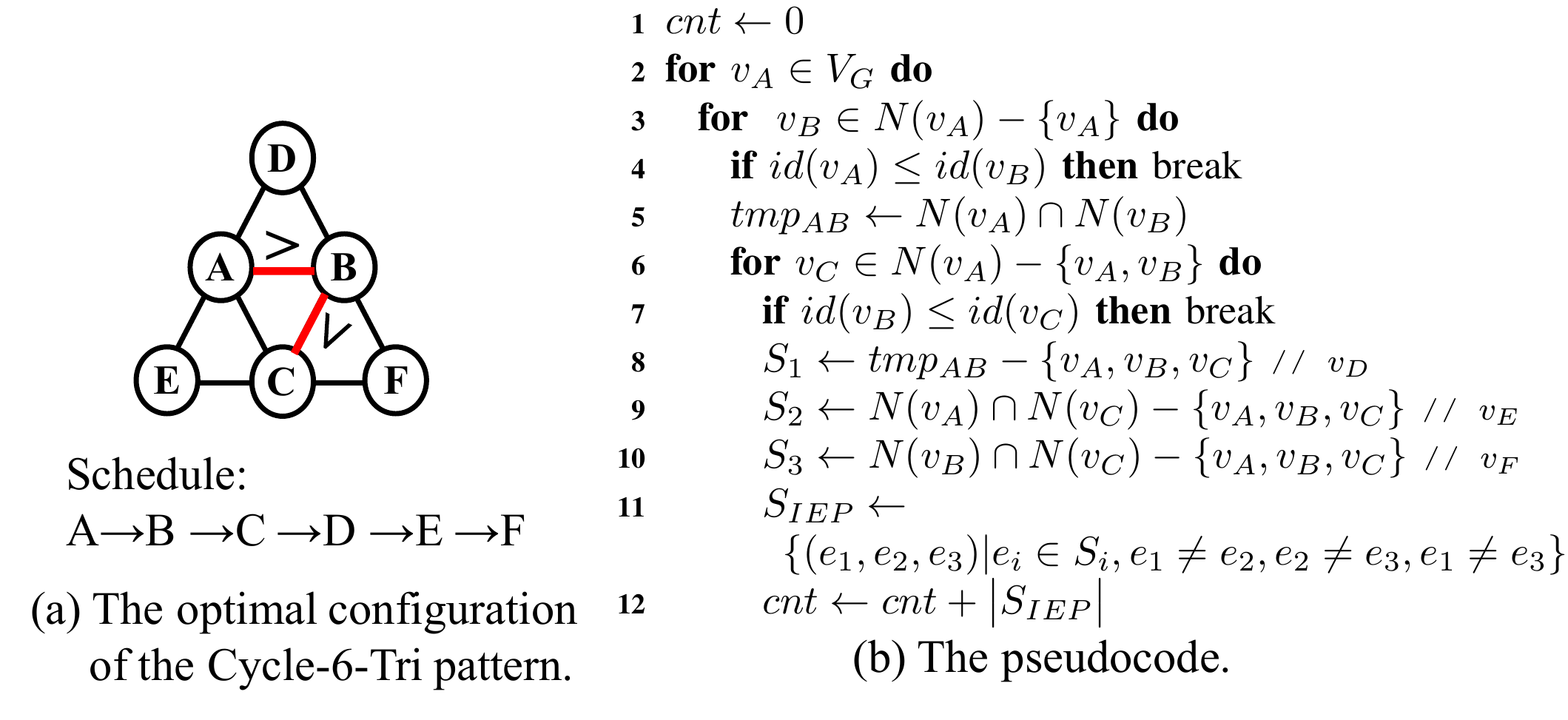}}
\setlength{\abovecaptionskip}{0.6em}
\setlength{\belowcaptionskip}{0.5em}
\caption{There are at most three vertices (D,E,F) in (a) that any two of them are not connected and therefore $k=3$. We use the Inclusion-Exclusion Principle to count the number of embeddings instead of searching $v_D$,$v_E$ and $v_F$ by three loops.}
\label{fig:cycle6-pseudocode}
\end{figure}

There are many graph mining problems, such as Clique Counting and Motif Counting,  which only need to collect the number of embeddings instead of listing all embeddings. This property provides us more optimization opportunities. According to the second phase in Section~\ref{schedule_generation}, there are usually no intersection operations in the innermost $k$ loops of the optimal configuration. Therefore, we leverage the technique of Inclusion-Exclusion Principle (IEP) instead of enumeration in the innermost $k$ loops to count embeddings. As shown in Figure~\ref{fig:cycle6-pseudocode}(b), $S_1$, $S_2$ and $S_3$ are the candidate sets of $D$, $E$ and $F$. Instead of using $v_D$, $v_E$ and $v_F$ to traverse through $S_1$, $S_2$ and $S_3$ respectively, we calculate the cardinality of $S_{IEP}$ in line 11 to count the number embeddings.

Without loss of generality, let $S_i$ be the candidate set of the $i$th vertex among the $k$ vertices. 
To count the number of embeddings, we need to calculate the cardinality of the set $S_{IEP}$, where 
\begin{subequations}
\nonumber
\begin{align}
    S_{IEP} = \;\; &S_1 \times S_2 \times \cdots \times S_k \\
              &- \left\{(e_1,e_2,\cdots,e_k) | \exists 1 \le i, j \le k \, \operatorname{ s.t. } \, e_i = e_j\right\}
\end{align}
\end{subequations}

We define 
\begin{equation}\nonumber
    A_{i,j} = \left\{(e_1,e_2,\cdots,e_k)| \forall 1\le l \le k, e_l\in S_l, e_i=e_j \right\}.
\end{equation}
For the convenience of description, we only consider $A_{i,j}$ which satisfies $1 \leq i < j \leq k$ in this Section. According to the Inclusion-Exclusion Principle, we have 
\vspace{0.5em}
\nonumber
{\allowdisplaybreaks[2]
\begin{align}
\big| S_{IEP} \big| &= \bigcap\limits_{i,j} \overline{A_{i,j}} \\ 
                    &= \big|U\big| - \Bigg|\bigcup\limits_{i,j}A_{i,j}\Bigg| \\
                    &= \prod\limits_{i=1}^{k}\big| S_i\big| - \sum\limits_{i,j}\big|A_{i,j}\big| \\
                    & + \sum\limits_{i_1,j_1} \; \sum\limits_{ i_1\times k+j_1 < i_2\times k + j_2}\big| A_{i_1,j_1} \bigcap A_{i_2,j_2} \big| - \cdots \\
                    & + (-1)^{\frac{k\times(k-1)}{2}} \Bigg| \bigcap\limits_{i,j}A_{i,j}\Bigg|,
\end{align}}
where $U$ is a universal set.

\begin{algorithm}[ht]
  \SetAlgoNoLine 
  \DontPrintSemicolon
  \caption{Cardinality Calculation}
  \label{alg:intersection_card}
  \KwIn{$k$}
  \KwIn{$i\_j\_pairs$}
  \KwIn{$S_1$,$S_2$,$\cdots$,$S_k$}
  \KwOut{$intersection\_card$: $\big|A_{i_1,j_1} \bigcap A_{i_2,j_2} \bigcap \cdots \bigcap A_{i_m,j_m}\big|$ where $(i_l,j_l)$ is the $l$th pair in $i\_j\_pairs$ and $m$ is the number of pairs.}
      $g \leftarrow$ an empty undirected graph with $k$ vertices \;
      \For{$pair \in i\_j\_pairs$}{$g$.add\_edge($pair.first$,$pair.second$)}
      $components \leftarrow$ partition g into connected components\;
      $intersection\_card \leftarrow 1$ \;
      \For{$comp \in components$}
      {
        $comp\_intersection \leftarrow$ universal set \;
        \For{$v \in comp$}{$comp\_intersection \leftarrow comp\_intersection \bigcap S_{v.id}$}
        $intersection\_card \leftarrow intersection\_card \times \big| comp\_intersection \big|$ \;
      }
      \KwRet $intersection\_card$
\end{algorithm}

To calculate $\big|A_{i_1,j_1} \bigcap A_{i_2,j_2} \bigcap \cdots \bigcap A_{i_m,j_m}\big|$, we use Algorithm \ref{alg:intersection_card}, which takes $(i_1,j_1),(i_2,j_2),\cdots,(i_m,j_m)$ as input and outputs the cardinality of the intersection. For each $(i,j)$ in the input pairs, we add an undirected edge between $i$ and $j$ in an initially empty undirected graph $g$ (lines 1$\sim$ 3). Next, we partition $g$ into connected components (line 4). Let $S_i$ be the set corresponding to the vertex $i$ in $g$. We perform intersection operations on the sets corresponding to each vertex in a connected component, and thus, there is an intersection ($comp\_intersection$) for every component (lines 6$\sim$9). We then get the answer by multiplying the cardinalities of the intersections (line 10). For example, to calculate $\big|A_{1,2}\bigcap A_{2,3} \cap A_{4,5}\big|$ when $k=6$, we need to add three edges $(1,2),(2,3)$, and $(4,5)$ into $g$. After partitioning $g$, there are three connected components: [1,2,3] , [4,5], and [6]. Then we have $\big|A_{1,2}\bigcap A_{2,3} \cap A_{4,5}\big| = \big|S_1 \bigcap S_2 \bigcap S_3 \big| \times \big|S_4 \bigcap S_5 \big| \times \big|S_6\big|$.

Note that we have not considered the restrictions during calculations with the Inclusion-Exclusion Principle. The restrictions in outer loops remain, but the restrictions in the innermost $k$ loops do not exist, which leads to overcounting of embeddings. By calling the function of $no\_conflict$ in Algorithm \ref{alg:restrictions_generation} for every permutation, we can get the number of permutations that cannot be eliminated by applying the remaining restrictions. Suppose that the number of permutations is $x$, then the correct number of embeddings is $\frac{ans_{IEP}}{x}$, where $ans_{IEP}$ is the result we calculate by counting with the Inclusion-Exclusion Principle.

\subsection{Distributed Implementation} \label{implementation}

To further improve the searching performance, we implement a distributed version of the pattern matching algorithm. We keep the whole data graph in the main memory and replicate it on each computing node. Although cross-node access to graph data can be avoided to reduce the overhead of network communication by keeping a complete copy of the data graph on each node, the size of a data graph that \systemName{} can handle is limited by memory. To handle larger graphs, distributed pattern matching faces the challenge of memory-aware graph partitioning and network load balancing, and we will solve these challenges in our future work. In the distributed implementation of \systemName{}, we mainly focus on the 
workload imbalance caused by task partitioning.

\systemName{} stores graphs in the compressed sparse row (CSR) format, that is, the neighborhood of a vertex is sorted and continuous in memory. Therefore, the intersection operation of two sets can be efficiently implemented with the time complexity of $O(n+m)$, where $n$ and $m$ are the cardinalities of the two sets respectively, and the intersection is naturally sorted.

Since the vertex degrees often follow a power-law distribution in real-world graphs, \systemName{} utilizes a fine-grained task partitioning technique to deal with the workload imbalance problem. In \systemName{}, there is a master thread responsible for partitioning and distributing tasks. The master thread executes the outer loops and packs the values of the outer loops into a task. Worker threads need to unpack tasks and continue executing the remaining inner loops. For example, in the case of the House pattern (Figure \ref{fig:pseudocode_performance}), suppose the master thread is executing the outmost two loops, and the current values of $v_A$ and $v_B$ are $x$ and $y$ respectively. Then the master thread sends the task of $(x,y)$ to a worker thread, and the worker thread will find all embeddings that satisfy $v_A=x$ and $v_B=y$. The number of outer loops executed by the master thread depends on the complexity of the pattern. For relatively simple patterns like the Triangle pattern, the master thread can balance the workload by executing only one loop (i.e., the outmost loop).

\systemName{} uses a work-stealing algorithm to schedule computations. There is a communication thread that maintains a task queue on each node. When the number of tasks in the task queue is less than a threshold, the communication thread uses asynchronous communication primitives of MPI to steal tasks from other nodes and add them to its queue. When a worker thread runs out of tasks, it takes one or more tasks from the task queue of its node.

\begin{figure}[b]
\centerline{\includegraphics[width=0.48\textwidth]{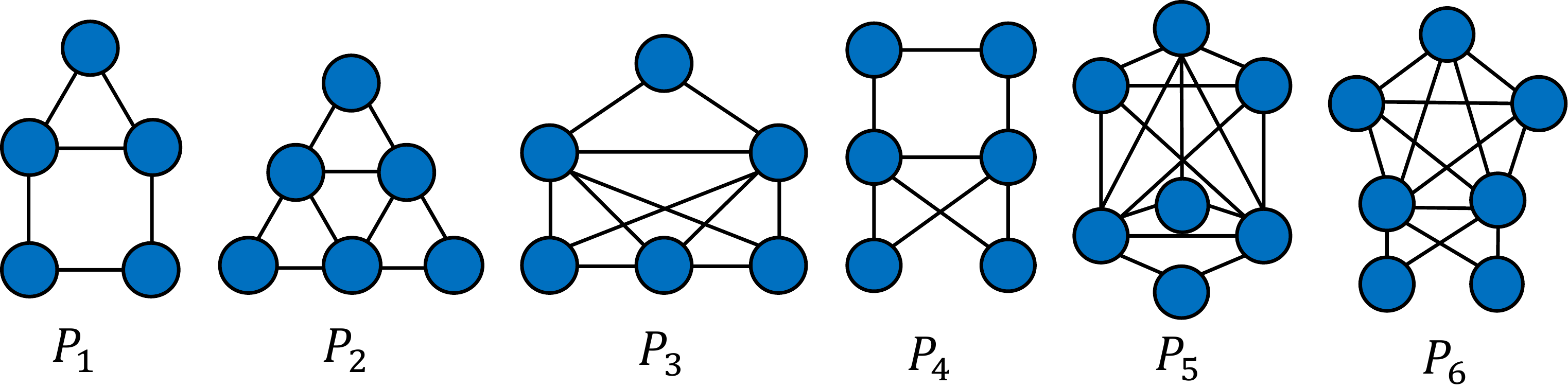}}
\caption{Patterns used in the evaluation.}
\label{fig:patterns}
\end{figure}

\begin{figure*}[htbp]
\centerline{\includegraphics[width=0.95\textwidth]{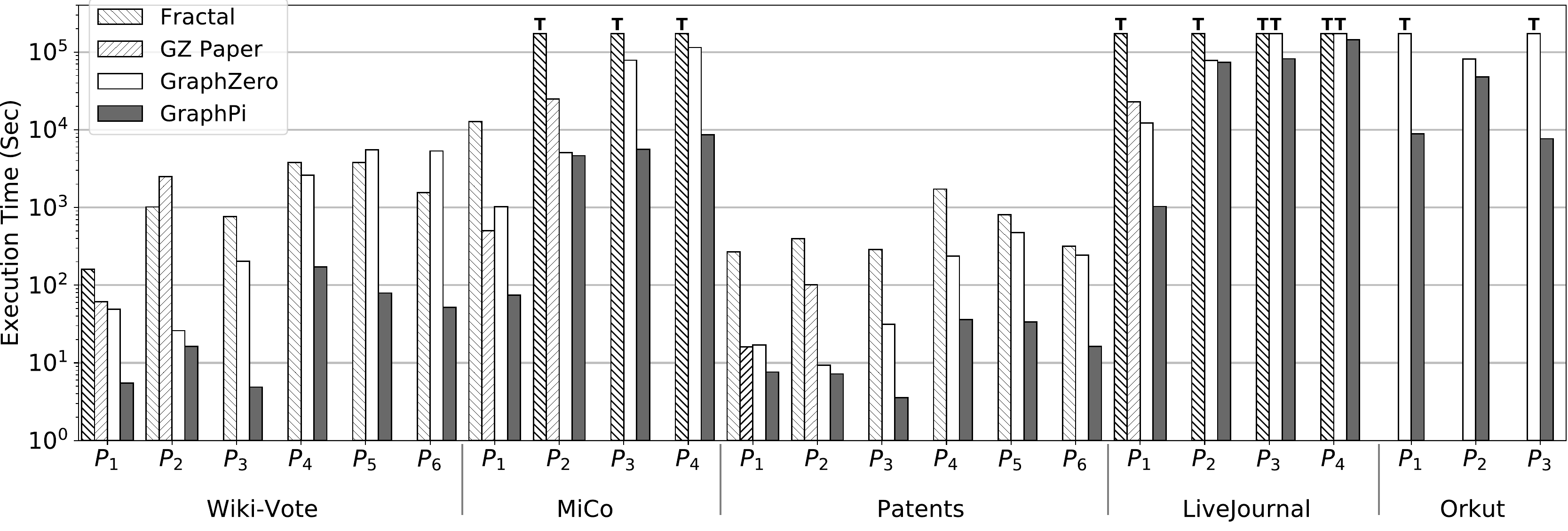}}
\caption{Overall performance of \systemName{}, GraphZero, and Fractal. ``T" means the execution time exceeds 48 hours.}
\label{fig:overall_performance}
\end{figure*}

\begin{figure}[b]
\centerline{\includegraphics[width=0.48\textwidth]{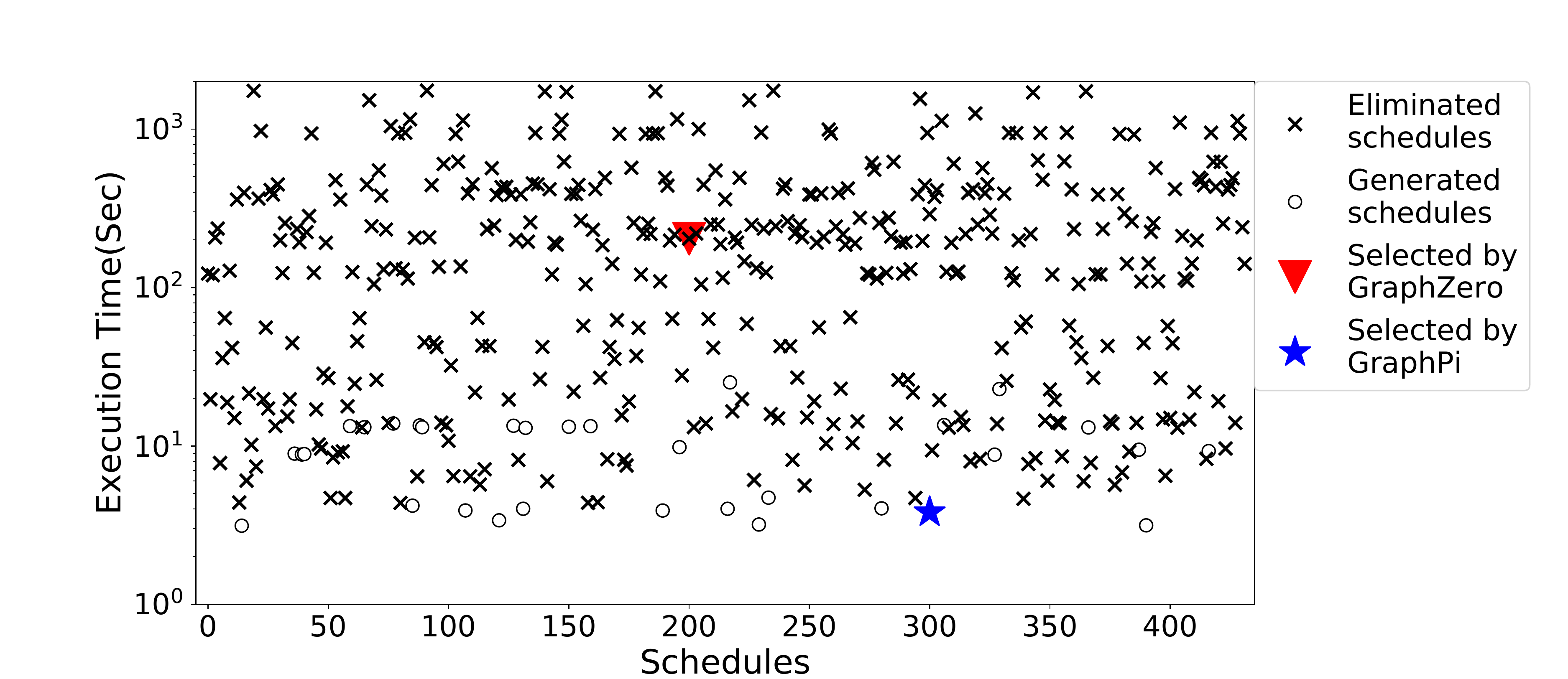}}
\caption{Performance of different schedules of $P_3$ on Wiki-Vote.}
\label{fig:schedules_performance}
\end{figure}
\section{Evaluation} \label{evaluation}

\subsection{Methodology} \label{methodology}

\para{Platforms} Our evaluation is performed on Tianhe-2A supercomputer. Each node has 2 12-core Intel Xeon E5-2692 (v2) processors (hyper-threading disabled) and 64GB of memory. We have implemented \systemName{} with an OpenMP/MPI hybrid programming model. On each node, we run 1 MPI process with 24 OpenMP threads. Tianhe-2A uses a customized high-speed interconnection network. Each node runs Kylin Linux with Linux kernel version 3.10.0 and gcc version 4.8.5. All programs are compiled with -O3 optimization option.

\begin{table}[htp!]
\caption{Graph datasets.} \label{datasets}
\small
\begin{tabular}{l|r|r|l}

\textbf{Graphs} & \textbf{\#Vertices} & \textbf{\#Edges} & \textbf{Description} \\ 
\hline
\hline
Wiki-Vote\cite{leskovec2010predicting} & 7.1K & 100.8K & Wiki Editor Voting\\
MiCo\cite{elseidy2014grami} & 96.6K & 1.1M & Co-authorship\\
Patents\cite{leskovec2005graphs} & 3.8M & 16.5M & US Patents\\
LiveJournal\cite{backstrom2006group} & 4.0M & 34.7M & Social network\\
Orkut\cite{yang2015defining} & 3.1M & 117.2M & Social network\\
Twitter\cite{kwak2010twitter} & 41.7M & 1.2B & Social network \\

\end{tabular}
\end{table}

\para{Datasets} We use 6 real-world graphs as shown in Table~\ref{datasets}. These graphs are also used in GraphZero~\cite{mawhirter2019graphzero}. The numbers of vertices and edges range from 7.1 thousand to 41.7 million, and 100.8 thousand to 1.2 billion, respectively. 
The largest graph (Twitter) is only used for scalability experiments.

\para{Patterns} We use six patterns as shown in Figure~\ref{fig:patterns}, and the first two of them are also used in GraphZero. We use four additional patterns since the patterns used in GraphZero are relatively simple, and it is trivial to predict the performance of different schedules for them.

\para{Comparison} We evaluate \systemName{}'s performance against GraphZero~\cite{mawhirter2019graphzero} and Fractal~\cite{dias2019fractal}, the state-of-the-art single-machine pattern matching systems. GraphZero is an upgraded version of AutoMine~\cite{mawhirter2019automine}, and it outperforms AutoMine by up to 40$\times$. Fractal is a JVM-based system, and it outperforms several JVM-based specialized algorithms (MRSUB~\cite{shahrivari2015distributed}, SEED~\cite{lai2016scalable} and QKCount~\cite{finocchi2015clique}) and general-purpose systems (Arabesque~\cite{teixeira2015arabesque} and GraphFrames~\cite{dave2016graphframes}) by orders of magnitudes. Since GraphZero is not released, we reproduce all the algorithms described in GraphZero and also compare its performance with the results reported in their paper. In the following experiments, we use ``GraphZero" to denote the performance of the reproduced version of GraphZero and use ``GZ Paper" to denote the performance results reported in GraphZero's paper. Since the definition of pattern matching in AutoMine and GraphZero is different from other systems, we made some minor modifications in the reproduced version to make its results consistent with those of other systems. When comparing with GraphZero and Fractal, \systemName{} runs on a single node and does not use the optimization with Inclusion-Exclusion Principle (IEP). The time reported in our evaluation does not include the graph loading time or the program preprocessing and compiling time.

To guarantee the correctness of \systemName{}, we compare GraphPi's results with those of Fractal and (the reproduced version of) GraphZero. The results show that the numbers of embeddings obtained by three systems are the same.

\subsection{Overall Performance} \label{overall_performance}
\begin{figure*}[htp]
\centerline{\includegraphics[width=0.95\textwidth]{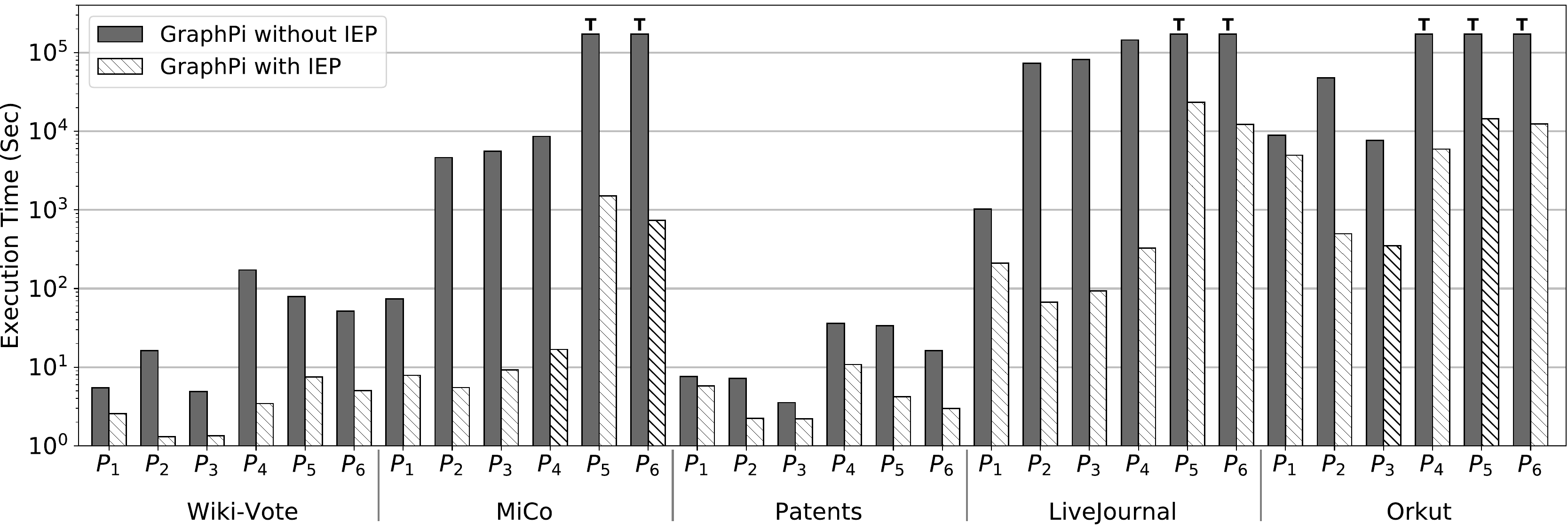}}
\caption{The performance of counting the number of embeddings with and without the Inclusion-Exclusion Principle (IEP). ``T" means the execution time exceeds 48 hours.
}
\label{fig:IEP_performance}
\end{figure*}
We perform experiments with \systemName{}, GraphZero, and Fractal to find all embeddings of the 6 patterns on 5 real-world graphs. Figure~\ref{fig:overall_performance} compares the performance of \systemName{} with GraphZero and Fractal in a log scale, and the performance results reported in GraphZero's paper are also shown on it. We only list the workloads that can be done within 48 hours. On average, \systemName{} outperforms GraphZero by 9.7$\times$, 1.4$\times$, 26.0$\times$, 11.7$\times$, 42.5$\times$, and 60.3$\times$ respectively for 6 patterns on different graphs. We get the highest speedup of 105$\times$ for $P_6$ on Wiki-Vote. In general, much higher speedups can be obtained on a larger graph, but GraphZero cannot finish the searching within 48 hours on large graphs. Since Fractal runs out of memory on Orkut, we only list the performance of Fractal on the other 4 graphs. On average, \systemName{} outperforms Fractal by 83.6$\times$, 64.9$\times$, 154.3$\times$, 35.5$\times$, 36.5$\times$, and 25.7$\times$ respectively for 6 patterns on different graphs. The speedup mainly comes from the optimal configuration used in the pattern matching algorithm automatically generated by \systemName{}. We will show more detailed results of our systems below. 

\subsection{Breakdown Analysis} 
\label{breakdown_analysis}

\para{Restriction Set Generation} To eliminate redundant computation caused by automorphisms, a set of restrictions need to be applied in pattern matching. There are many different sets of restrictions for one pattern, but their performance varies significantly. For a given schedule of a pattern, \systemName{} generates different sets of restrictions and selects the best one based on our performance prediction model. In contrast, GraphZero can only generate one set of restrictions, and sometimes it may achieve sub-optimal results. 

\begin{table}[htbp]
\caption{Speedup obtained with the better set of restrictions selected by \systemName{}.}
\center
\newcommand{\tabincell}[2]{\begin{tabular}{@{}#1@{}}#2\end{tabular}}
\begin{tabular}{c|c|c|c}
  
  \textbf{Graph} & \textbf{Pattern} & \tabincell{c}{\textbf{Average}\\ \textbf{Speedup} }& \tabincell{c}{\textbf{Maximum}\\ \textbf{Speedup} }\\
  \hline
  \hline
   & $P_1$ & 1.94 & 2.52\\
  Wiki-Vote & $P_2$ & 1.71 & 4.10\\
   & $P_4$ & 1.60 & 2.39\\
  \hline
   & $P_1$ & 2.02 & 5.08\\
  Patents & $P_2$ & 1.65 & 6.65\\
   & $P_4$ & 2.46 & 7.82\\
  
\end{tabular}
\label{tab:restriction_speedup}
\end{table}

We run all schedules of $P_1$, $P_2$, and $P_4$ with \systemName{} and GraphZero on Wiki-Vote and Patents. The restriction sets selected by \systemName{} and GraphZero are different in some schedules, which leads to varied performance. For these schedules, we compare the performance of \systemName{} with that of GraphZero. Table~\ref{tab:restriction_speedup} shows the average speedup and maximum speedup obtained with \systemName{} over GraphZero. \systemName{} achieves up to 7.8$\times$ speedup than GraphZero for the same schedule. 
The speedup mainly comes from a better set of restrictions selected by \systemName{}. These results also demonstrate that generating multiple sets of restrictions for a schedule in \systemName{} provides more choices and optimization opportunities for the pattern matching algorithm.

\para{Schedule Generation and Selection} The selection of schedules also has a significant influence on the performance of the pattern matching. We perform experiments with schedules generated by \systemName{} and GraphZero for each pattern on Wiki-Vote and Patents. To avoid the influence of different sets of restrictions on performance, we use the same restriction generation algorithm proposed by GraphZero for both \systemName{} and GraphZero in this experiment. On average, \systemName{} outperforms GraphZero by 25.6$\times$ for 6 patterns on Wiki-Vote and Patents. The speedup comes from \systemName{}'s computation-avoid schedule generator and accurate performance model.

Figure~\ref{fig:schedules_performance} shows the performance of different schedules of $P_3$ on Wiki-Vote, including two final schedules selected by \systemName{} (denoted by the blue star) and GraphZero (denoted by the red triangle). The ``$\circ$" marker denotes the schedules generated by \systemName{}'s 2-phase computation-avoid schedule generator, and the ``$\times$" marker denotes the schedules eliminated by the generator. We can see that most schedules with low performance (including the one selected by GraphZero) are eliminated, which shows the benefit of our 2-phase schedule generator in accurately generating efficient schedules. Among all efficient schedules (denoted by ``$\circ$"), the oracle schedule is 8.0$\times$ faster than the slowest schedule. This is because our performance prediction model can further select the optimal schedule (blue star marker) from these efficient schedules, which is only 22\% slower than the oracle schedule.

\begin{figure}[h]
\centerline{\includegraphics[width=0.48\textwidth]{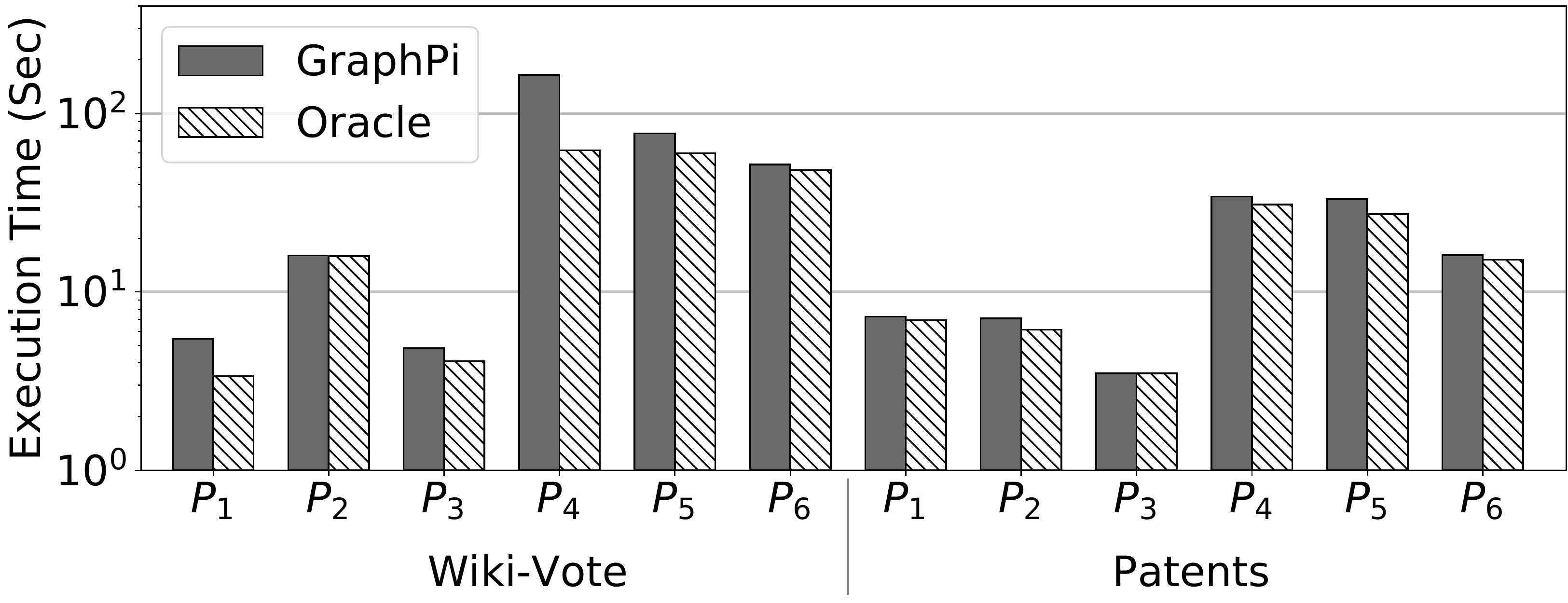}}
\caption{The accuracy of \systemName{}'s performance prediction model.}
\label{fig:performance_model_accuracy}
\end{figure}

\para{Accuracy of Performance Prediction Model} To further demonstrate the accuracy of \systemName{}'s performance prediction model, we perform experiments with all schedules generated by \systemName{} for each pattern on Wiki-Vote and Patents. Figure~\ref{fig:performance_model_accuracy} shows the performance of the schedules selected by \systemName{} and the oracle schedules. On average, schedules selected by GraphPi are only 32\% slower than the oracle schedules. There is a performance gap between the selected schedule and the oracle schedule of $P_4$ on Wiki-Vote, this is because the prediction of the number of rectangles (i.e., the subpattern formed by the top 4 vertices of $P_4$) on Wiki-Vote is not accurate enough due to the insufficient structural information we leverage (only the numbers of vertices, edges and triangles). To achieve more accurate prediction, we need to use more structural information of data graphs in the performance prediction model.

\subsection{Counting Embeddings with IEP} 



We also do optimization with the Inclusion-Exclusion Principle (IEP) in \systemName{} instead of enumeration to count the number of embeddings. We evaluate the performance of \systemName{} when enabling IEP or not for each pattern. We use the same configuration selected by \systemName{}'s performance model in experiments, that is, we avoid the influence of schedules and sets of restrictions on performance. As shown in Figure~\ref{fig:IEP_performance}, counting with IEP outperforms that without IEP by 4.3$\times$, 457.8$\times$, 320.5$\times$, 265.5$\times$, 11.1$\times$, and 10.1$\times$ respectively for 6 patterns on different graphs on average. We obtain the highest speedup of 1110.5$\times$ for $P_2$ on LiveJournal.

\begin{figure}[h]
  \centering%
  \subcaptionbox{Orkut}
    {\includegraphics[width=0.23\textwidth]{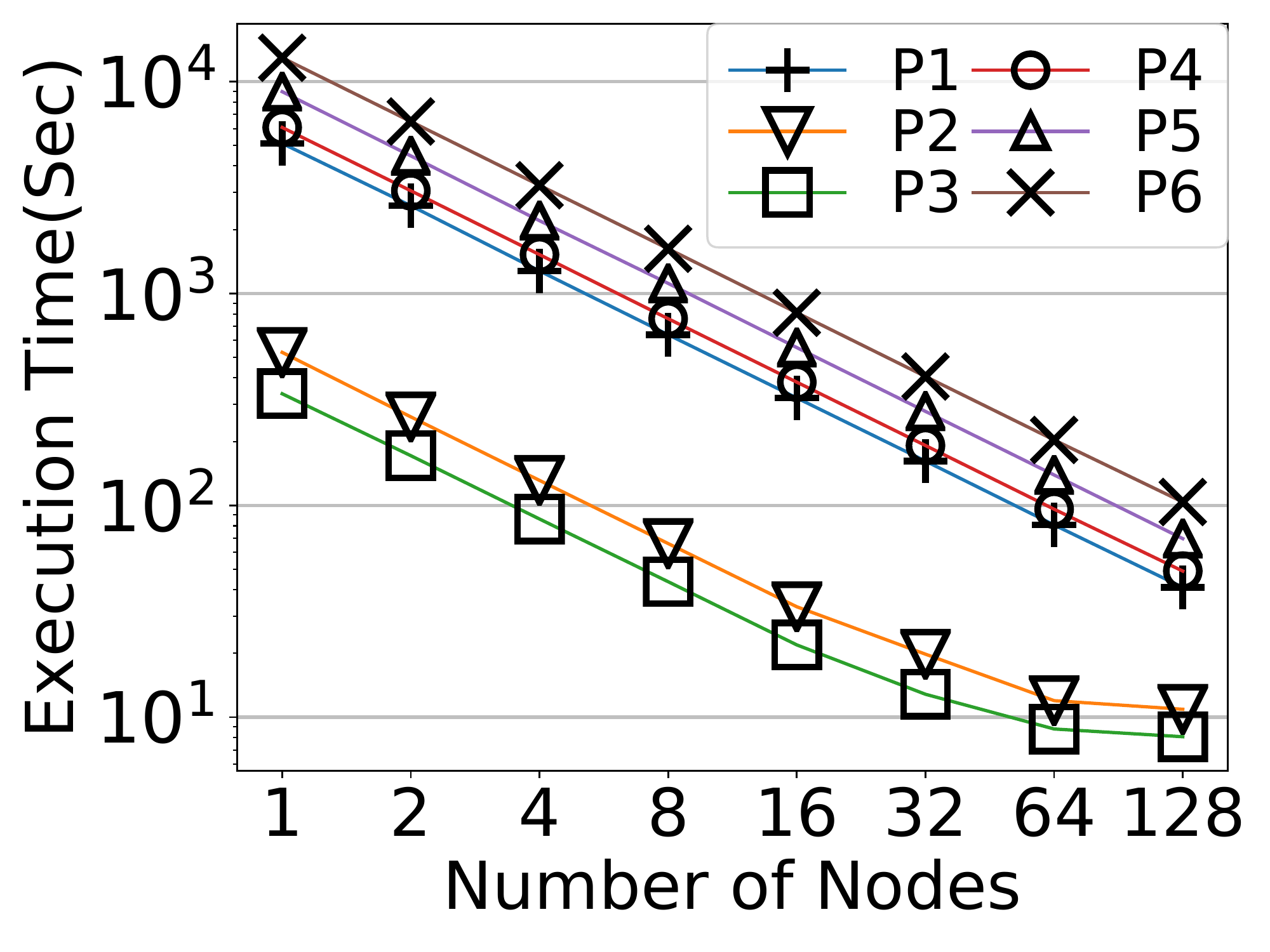}\vspace{-0.6em}}%
  \hspace{0.5em}%
  \subcaptionbox{Twitter}
      {\includegraphics[width=0.23\textwidth]{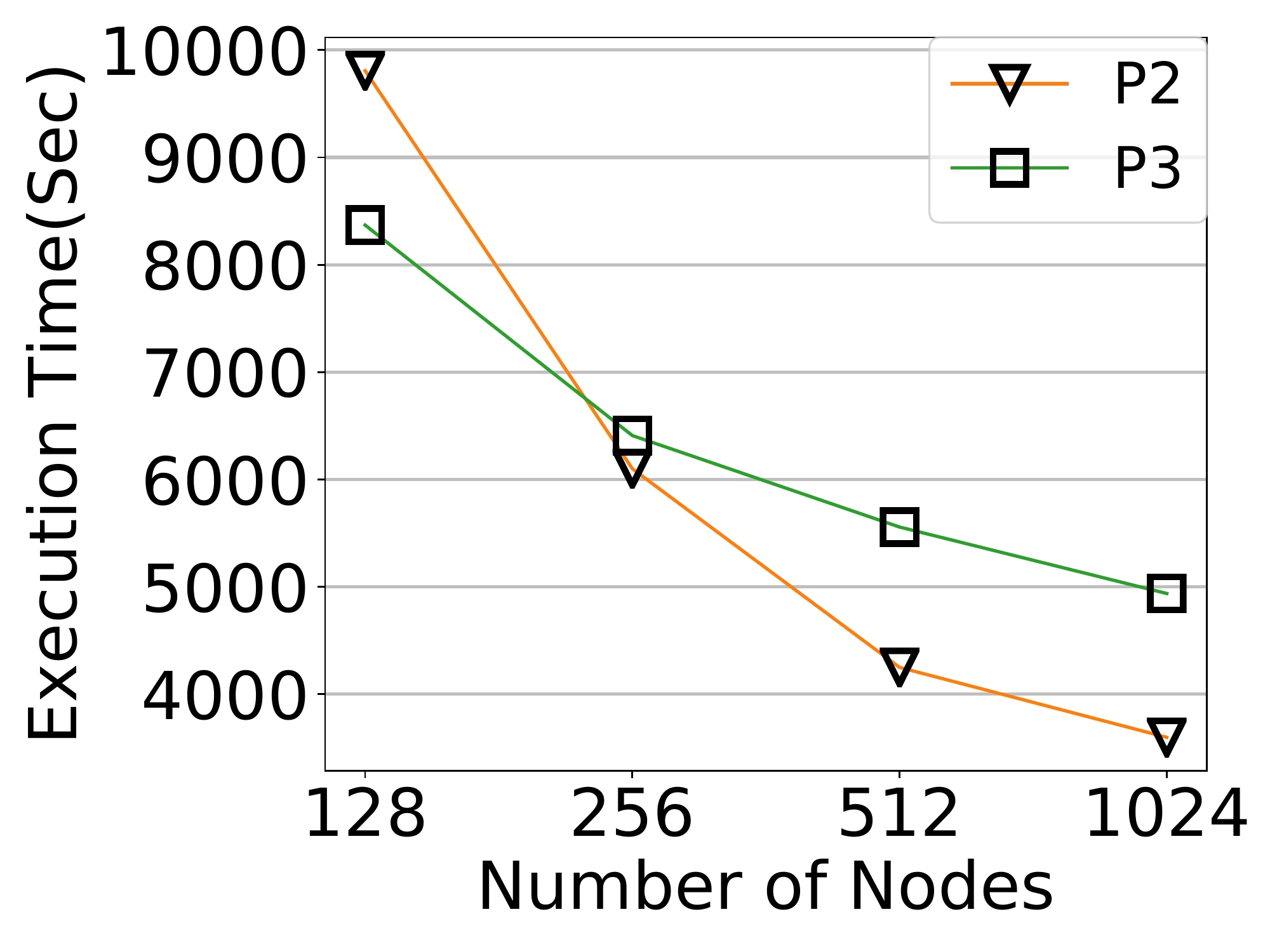}\vspace{-0.6em}}
  \caption{Scalability of \systemName{}.}
  \label{fig:scalability}
\end{figure}

\subsection{Scalability of Distributed Version}
In this subsection, we evaluate the scalability of \systemName{} system with up to 1,024 nodes (24,576 processor cores) on Tianhe-2A supercomputer. Figure \ref{fig:scalability} shows the speedup obtained with the increasing number of nodes on Orkut and Twitter. \systemName{} obtains a near-linear speedup with 128 nodes when running $P_1$, $P_4$, $P_5$, and $P_6$ on Orkut. $P_2$ and $P_3$ do not scale very well on Orkut because their execution times are very short (10.9 seconds and 8.1 seconds respectively). For the experiment on Twitter, since the execution times of other patterns with 128 nodes exceed 24 hours, we only evaluate $P_2$ and $P_3$ on 128$\sim$1024 nodes. 
\systemName{} does not get linear speedups for $P_2$ and $P_3$ on Twitter due to load imbalance. In the future, we plan to implement much fine-grained subtask partitioning in \systemName{} to solve this problem.

\begin{table}[htbp]
\caption{The overhead of preprocessing and code generation in \systemName{} for each pattern.} \label{tab:overhead}
\center
\begin{tabular}{c|c}
  
  \textbf{Pattern} & \textbf{Overhead (second)} \\
  \hline
  \hline
  $P_1$ & 0.008 \\
  $P_2$ & 0.07 \\
  $P_3$ & 0.04 \\
  $P_4$ & 0.07 \\
  $P_5$ & 1.88 \\
  $P_6$ & 2.53 \\
  
\end{tabular}
\end{table}

\subsection{Overhead of Preprocessing and Code Generation}

The time reported in previous experiments does not include the preprocessing (i.e., configuration generation and performance prediction) and code generation time. In this subsection, we evaluate the overhead of preprocessing and code generation in \systemName{}. As can be seen from the preprocessing algorithm, the overhead is only related to the structure of a pattern but not to the input data graph. Table~\ref{tab:overhead} shows the overhead of preprocessing and code generation ranging from 8 milliseconds to 2.53 seconds. Compared with the execution time of the pattern matching that may take several minutes or even several hours, the overhead can be ignored.
\vspace{1.0em}
\section{Related Work}

\para{General-Purpose Graph Mining Systems} Arabesque~\cite{teixeira2015arabesque} is the first distributed graph mining system that provides a high-level abstraction and a flexible programming model. It maintains some intermediate data of subgraphs and generates all pattern instances by appending edges to candidate subgraphs and filtering the newly generated candidates using the user-defined filter and process functions. G-thinker~\cite{ yan2017g} provides an intuitive graph-exploration API for implementing various graph mining algorithms and an efficient runtime engine.

G-Miner~\cite{chen2018g} models the processing of a graph mining job as an independent task and streamlines task processing with a novel design. RStream~\cite{wang2018rstream} is a single-machine system which implements relational algebra efficiently with tuple streaming. To support scalable graph mining, RStream uses out-of-core processing to leverage disk support to store intermediate data.

\para{Graph Pattern Matching Systems} Although general-purpose graph mining systems provide flexible programming models to express complex graph mining algorithms, their performance is relatively poor. Specialized pattern matching systems have been proposed~\cite{cordella2004sub, han2013turboiso, serafini2017qfrag, ren2015exploiting, bhattarai2019ceci, reza2018prunejuice}. Automine~\cite{mawhirter2019automine} is built upon a set-based representation and uses compilation techniques to generate efficient pattern matching code. However, due to the inherent symmetry in the structural patterns, Automine’s algorithm causes substantial computation redundancy. Based on Automine, GraphZero~\cite{mawhirter2019graphzero} provides an algorithm based on group theory to break the inherent symmetry in patterns and eliminate redundant computation. Peregrine~\cite{jamshidi2020peregrine} is another DFS-based system which provides a pattern-based programming model and a workflow similar to GraphZero. Peregrine also has a schedule generation module. However, the schedule generated by Peregrine is only based on the pattern, without considering the distribution of data in different data graphs.

\para{Approximate Subgraph Counting} Some computation engines and systems have been designed to estimate an approximate number of embeddings~\cite{bordino2008mining, alon2008biomolecular, gonen2011counting}. ASAP~\cite{iyer2018asap} is the state-of-the-art one among them. ASAP is a distributed approximate computation engine for graph pattern matching. Based on the neighborhood sampling algorithm, ASAP samples embeddings from the stream of the edge set of a graph to do an approximate estimation. Although these approximate systems have good scalability, they cannot list all embeddings.
\section{Conclusion}
In this work, we propose \systemName{}, a high-performance distributed pattern matching system.
We design a 2-cycle based automorphism elimination algorithm and an accurate performance model to eliminate redundancy in pattern matching. 
When counting embeddings, we further propose using the Inclusion-Exclusion Principle to achieve significant improvement in performance.
Results show that \systemName{} outperforms the state-of-the-art pattern matching system by up to two orders of magnitude on a single node and can scale to 1,024 nodes.

\balance
\bibliographystyle{IEEEtran}
\bibliography{refs}

\end{document}